\documentclass[authoryear]{elsarticle}

\journal{Solar Energy}

\usepackage{graphicx}

\usepackage{amssymb,amsmath}
\usepackage{enumitem}
\usepackage{mathtools}
\usepackage{comment}
\usepackage{url}
\usepackage[hyperindex,pdftex]{hyperref}

\usepackage{comment}
\usepackage{algorithm,algorithmic}
\usepackage{placeins}

\usepackage{color}
\usepackage[dvipsnames]{xcolor}

\newtheorem{definition}{Definition}

\newtheorem{problem}{Problem}
\newtheorem{assumption}{Assumption}
\newtheorem{remark}{Remark}

\begin{document}

\begin{frontmatter}



\title{Nonlinear Temperature Regulation of Solar Collectors with a Fast Adaptive Polytopic LPV MPC Formulation}

\author[adutn]{Hugo A. Pipino}
\ead{hpipino@sanfrancisco.utn.edu.ar}
\author[adufsc]{Marcelo M. Morato}
\ead{marcelomnzm@gmail.com}
\author[adutn]{Emanuel Bernardi}
\ead{ebernardi@sanfrancisco.utn.edu.ar}
\author[adunl]{Eduardo J. Adam}
\ead{eadam@fiq.unl.edu.ar}
\author[adufsc]{Julio E. Normey-Rico}
\ead{julio.normey@ufsc.br}

\address[adutn]{Applied Control \& Embedded Systems - Research Group \emph{(AC\&ES-RG)},\\
Universidad Tecnol\'ogica Nacional, San Francisco, Argentina.}

\address[adufsc]{Renewable Energy Research Group \emph{(GPER)}, Departamento de Automa\c{c}\~ao e Sistemas \emph{(DAS)},\\ Universidade Federal de Santa Catarina, Florian\'opolis, Brazil.}

\address[adunl]{Facultad de Ingenier\'ia Qu\'imica, Universidad Nacional del Litoral, Santa Fe, Argentina.}



\begin{abstract}
Temperature control in solar collectors is a nonlinear problem: the dynamics of temperature rise vary according to the oil flowing through the collector and to the temperature gradient along the collector area. In this way, this work investigates the formulation of a Model Predictive Control (MPC) application developed within a Linear Parameter Varying (LPV) formalism, which serves as a model of the solar collector process. The proposed system is an adaptive MPC, developed with terminal set constraints and considering the scheduling polytope of the model. At each instant, two Quadratic Programming (QPs) programs are solved: the first considers a backward horizon of $N$ steps to find a virtual model-process tuning variable that defines the best LTI prediction model, considering the vertices of the polytopic system; then, the second QP uses this LTI model to optimize performances along a forward horizon of $N$ steps. The paper ends with a realistic solar collector simulation results, comparing the proposed MPC to other techniques from the literature (linear MPC and robust tube-MPC). Discussions regarding the results, the design procedure and the computational effort for the three methods are presented. It is shown how the proposed MPC design is able to outrank these other standard methods in terms of reference tracking and disturbance rejection.
\end{abstract}


\end{frontmatter}


\section{Introduction}
Efficient energy generation is one of essential tasks for ambitious sustainability goals. Recent academic research has given focus to the use of renewable-based systems to power and diversify energy matrices. Their integration is indeed a good alternative to avoid greenhouse emissions and environmental impact \cite{shafiee2009will,morato2018IJEPES}.

Accordingly, the use of solar energy has significantly increases during the last decades; various kinds of applications are today available, such as photovoltaic panels, solar-thermal collectors and others \cite{camacho2012control}. Solar energy is widespread, used in many countries, for different purposes \cite{badescu2007optimal, zambrano2008model, powell2012modeling, lima2016temperature, costa2016optimal}.

One of the major technological trends of solar energy is to use the radiance power to heat fluids for industrial and residential purposes: in distillation units for fresh water production \cite{alarcon2005design}, in bio-reactors to produce bio-sources (biomas, biogas) \cite{fernandez2012dynamic}, among many other applications; low-temperature solar-thermal plants are widely used \cite{leblanc2010low,bujedo2011experimental,marc2012assessing, sharma2015automatic}.  These solar-thermal (ST) collectors, with the heat coming from the solar radiance, must be have their output temperature regulated according to the application, since these hot fluids outlets are fed into the main stage of the cascaded systems. The temperature of the heated fluid should allow the correct operation of the main stages, which constitutes an important and complex control problem, since nonlinear dynamics and partial differential equations are involved \cite{morato2019SBAI}. 

Model Predictive Control (MPC) is a widely accepted control toolkit, which is generally understood as the range of optimization methods that inherently embed prediction models for the controlled process output \cite{camacho2013model}. Moreover, these strategies find an optimal policy by minimizing a cost function over a receding horizon, analytically including state, input and output constraints \cite{normey2007control}. This cost function includes performance goals, such as reference tracking and disturbance rejection. MPC is a solid candidate to control these modern ST plants \cite{lima2016temperature, DFREJO2020190}, and have been accordingly applied for many ST applications: for ST air-conditioning systems \cite{camacho2007surveyI, camacho2007surveyII}, for solar furnaces \cite{beschi2011constrained}, for swimming pools heating systems \cite{Delgadomarin2019use}, for steam-turbines to generate electricity \cite{galvez2009nonlinear}, and many others \cite{torrico2010robust, saade2014model, rahmani2015nonlinear, alsharkawi2016dual, Sanchez2019adaptive, Morato2020qLPVMPC, bella2020robust}.

All these previously-cited papers can be arranged into two major groups \cite{camacho2007surveyII}: (\textit{i}) those that linearize or simplify the nonlinearities of the ST heating process \cite{torrico2010robust, beschi2011constrained, lima2016temperature, Morato2020qLPVMPC}, which achieve (sometimes very decent, but) sub-optimal control results; and (\textit{ii}) those that opt to include the nonlinearities into the optimization problem or treat them robustly (as uncertainty blocks) \cite{galvez2009nonlinear, rahmani2015nonlinear, burger2018algorithm}. 

The original MPC algorithms were mainly attached to the scope of linear time-invariant (LTI) models, using 
state-space formulations. The solution of linear MPC is found by solving a constrained Quadratic Programming (QP) program. These formulations are the ones used for the first set of papers (\textit{i}). When considering the nonlinearities of the ST process, the prediction of the variables along the horizon becomes an incipient issue, since the inclusion of nonlinear predictions is not at all trivial and much increases the complexity of optimization problem \cite{allgower2012nonlinear}, making the algorithm difficult to run in real-time. These nonlinear MPC (NMPC) algorithms, if sought to be really implementable (fast enough) must be adapted, by reducing complexity and resorting to some sub-optimality, as it is done with Real-Time iteration methods, such as ACADO \cite{houska2011auto}, and gradient-based methods, such as GRAMPC \cite{kapernick2014gradient}.

It must be remarked that, in parallel to the growth of predictive control applications, literature became very rich on design methods for Linear Parameter Varying (LPV) systems \cite{mohammadpour2012control,sename2013robust}, although LPV models for ST system are rather scarce. Such systems are nonlinear ones that depend on a vector of known, bounded scheduling parameter, denoted as $\rho$. Thanks to Linear Differential Inclusion (LDI), nonlinear systems can be represented within an LPV setting, with simple (LTI alike) mathematical frameworks. As previously discussed in the literature \cite{cisneros2017fast}, the use of LPV models for nonlinear systems enables real-time, fast applications.

With respect to these aforementioned works, it becomes evident that fast MPC methods for ST systems are lacking. Moreover, comparison in terms of numerical efficiency and achieved performances between the two sets of works (\textit{i}) and (\textit{ii}) is also lacking. Therefore, the main motivation of this paper is to propose a novel, fast MPC scheme for the nonlinear temperature control of modern ST units. 

The proposed scheme is an adaptive control method that determines the optimal control policy through two consecutive QPs. The first QP works much like a Moving Horizon Estimator (MHE) \cite{rawlings2006particle, kuhl2011realparameter}, which uses available data from a backward horizon of $N$ steps and minimizes the difference from the data and a polytopic LPV model with a fixed virtual tuning variable; assuming this variable remains constant throughout the following $N$ steps, a regular LTI MPC problem is solved in the second QP. The proposed adaptive MPC (AMPC) method also includes terminal ingredients (stage cost and set constraint) to ensure stabilization despite the model simplifications. Previously, in \cite{morato2019LPVS,suboptimalMate2019}, it has been shown that sub-optimal QP design can be recursively feasible with these set-based constraints.

Thus, the contributions presented in this paper are summarized:
\begin{itemize}
\item Considering a polytopic LPV model for a ST system, an adaptive set-based MPC design procedure (Section \ref{sec3}) is formalized. As explained, this algorithm is based on two consecutive QPs that take into account the scheduling polytope to find the best LTI prediction for the next $N$ steps (horizon). The set-based tools are included to ensure stabilization.
\item For comparison purposes, a robust MPC with trajectory tubes is recalled for the case of LPV models (Section \ref{sec4}). The use of MPC algorithms based on projection tubes \cite{Mayne:2,rakovic2012parameterized,limon2010robust} has been previously shown to yield good results for the LPV case \cite{hanema2017stabilizingother, hanema2017stabilizing}.
\item Then, via high-fidelity numerical simulation, the effectiveness of the proposed AMPC tool is compared to a linearization-based MPC and to a tube MPC. These results are demonstrated for the nonlinear outlet temperature regulation problem (Section \ref{sec5}). Discussions are presented in order to evaluate the achieved performances and implementation drawbacks of each method. Considering the results, the methods are also compared in terms of number of constraints, amount of offline preparation, performances and online complexity. 
\end{itemize}

Regarding organization, the rest of the paper is structured as follows. In Section \ref{secSTunits}, the modern ST heating systems under investigation are formally presented, in terms of models, constraints and performance goals. Then, in Section \ref{sec2}, the predictive control problem for LPV models for regulation is defined, making evident how the evolution of $\rho$ becomes a computational issue, since: \textit{i)} it is (\textit{a priori}) unknown; and \textit{ii)} it transforms the optimization procedure into a nonlinear one. As mentioned, Sections \ref{sec3},\ref{sec4} and \ref{sec5} present, respectively, the proposed controller, a tube-based MPC approach and the comparison results in terms of simulations. The paper conclusions are drawn in Section \ref{seconc}.

Before the development of the paper, the following definitions are recalled:

\begin{definition}{Nonlinear Programming Problem}\\
Consider an arbitrary real-valued nonlinear function $f_c(x_c)$. A
Nonlinear Programming Problem (NP) finds the vector $x_c$ that minimizes
$f_c(x_c)$ subject to $f_i(x_c) \, \leq \, 0$, $f_e(x_c) \, = \, 0$
and $x_c \, \in \, \mathcal{X}_c$, where $f_i$ and $f_e$ are also nonlinear functions and $\mathcal{X}_c$ the admissible set.
\end{definition}

\begin{definition}{Quadratic Programming Problem}\\
A Quadratic Programming Problem (or simply Quadratic Problem, QP) is a linearly constrained mathematical optimization problem of a quadratic function. A QP is a particular type of nonlinear programming problems. The quadratic function may be defined with respect to several variables, all of which may be subject to linear constraints. Considering a vector $c \, \in \, \mathbb{R}^{n_c}$, a symmetric matrix $Q_c \, \in \, \mathbb{R}^{n_c \times n_c}$, a real matrix $A_{ineq} \, \in \, \mathbb{R}^{m_c \times n_c}$, a real matrix $A_{eq} \, \in \, \mathbb{R}^{m_c \times n_c}$, a vector $b_{ineq} \, \in \mathbb{R}^{m_c}$ and another vector $b_{eq} \,
\in \mathbb{R}^{m_c}$, the goal of a QP is to determine the vector $x_c \, \in \, \mathbb{R}^{n_c}$ that minimizes a regular quadratic function of form $\frac{1}{2}\left(x_c^TQ_xx_c + c^Tx_c\right)$ subject to constraints $A_{ineq}x_c \, \leq \, b_{ineq}$ and $A_{eq}x_c \, = \, b_{eq}$. The solution $x_c$ to this kind of problem is found by
many solvers seen in the literature, based on Interior Point algorithms, quadratic search, etc.
\end{definition}

\section{Temperature Control in Solar Collectors}
\label{secSTunits}

Adding renewable energy sources to power plants can be a good route to reduce greenhouse gas emissions and environmental impact. Anyhow, an inherent problem to be solved is how to integrate these energy sources without loosing efficiency and dispatchability of energy plants. 

As discussed in the literature \cite{camacho2012control}, current solar energy technologies are of two main kinds: photovoltaic systems, that directly covert solar radiance into electric energy, and solar-thermal systems, which usually generate heated fluid (or steam). The focus of this paper is the second category.

\begin{figure}[ht]
\centering
\includegraphics[width=\linewidth]{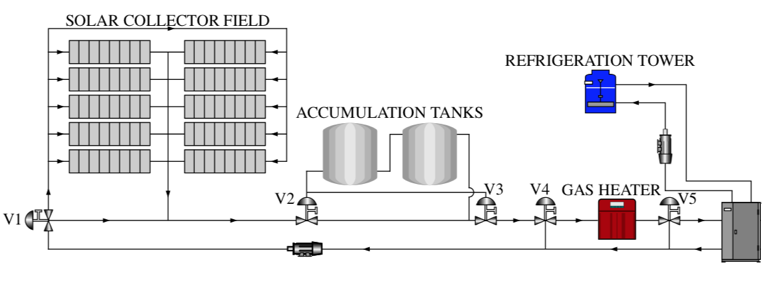}
\caption{\label{theCIESOL} Schematic Illustration of a Modern Solar-Thermal Collector System, comprising the solar collector field, accumulation tanks to store the heated fluid, a gas heater to further heat the liquid, if necessary, and a refrigeration tower.}
\end{figure}

Solar radiance is an intermittent energy source. When there occurs a cloudy period of the day, for instance, energy might be running low if no compensation strategy is considered. A practical solution for this matter, adopted in the majority of modern ST systems \cite{Dell:2001}, is to include accumulation tanks to store energy (hot fluid) while there is no process demand, and a complementary (auxiliary) energy source (say, for instance, a gas heater), that could be of use when there is no sun or (and) the accumulation tanks are not sufficient to meet the demand fully. A modern ST unit is usually a structure that integrates a solar-thermal collector field, some accumulation tanks and a gas heater. Of course, each subsystem has independent dynamics that influence strongly the total output dynamics. Nonetheless, in this paper, global coordination as well as the control of the tanks and gas heaters are regularly working\footnote{A global coordination algorithm for modern ST systems has been proposed recently \cite{de2019mixed}, using mixed logical MPC.}, the focus is solely in the temperature regulation of the ST collector panel itself. Figure \ref{theCIESOL} gives an illustration of such complete ST collectors.

\subsection{The CIESOL ST Plant}
Complete phenomenological models have previously been derived for ST collector fields \cite{pasamontes2013hybrid, gallego2013observer}, which according model-validation \cite{ampuno2019apparent} and parameter identification procedures \cite{morato2019SBAI}. In this work, the model is based on the \emph{CIESOL ST} plant, located in the
\emph{CIESOL-ARFR-ISOL} \emph{R}\&\emph{D} Centre of the University of
Almer\'ia, Spain. This testbed has a flat \emph{ST} collector, used to regulate the temperature of the inlet fluid to guarantee a certain heat demand.

Regarding these phenomenological models, they are based on the following assumptions:
\begin{itemize}
\item The fluid flow through the solar collector is incompressible, with uniform pressure along the field;
\item The heat transfer capacity of the collector plates is constant and denoted $C_m$; the density of these metal plates is also constant and denoted $\epsilon _f$;
\item The balance of energy equations assume a constant thermal loss coefficient $\nu$, with respect to the thermal energy that derives from the incident solar radiance;
\item The heat transfer coefficient of the absorver (external temperature to plates), denoted $h_0$, is constant, while the heat transfer coefficient of the fluid (fluid to plates), denoted $h_i(\cdot)$, varies positively according to the temperature of the plates.
\end{itemize}

Then, the following  partial-differential dynamics arise due to balance of energy equations, where $t$ represents the time variable and $s$ the space variable:
\begin{eqnarray}\label{modfinalcoletor1}
\xi_mC_mA_{e} \frac{dT_p}{dt}(t) &=&d_e\pi\nu I(t) - d_e\pi h_0(T_p(t) - T_e(t)) - d_i\pi h_i(T_p(t))(T_p(t) - T_f(t)) \, \text{,} \\\label{modfinalcoletor2}
\xi_fC_fA_{i} \frac{\partial T_{f}}{\partial t} (t,s) &=& -u(t)\epsilon_fC_f\frac{\partial T_f}{\partial s}(t,s) + d_i\pi h_i(T_p(t))(T_p(t) - T_f(t)) \, \text{.}
\end{eqnarray}

In these temperature gradient dynamics of Eqs. \eqref{modfinalcoletor1}-\eqref{modfinalcoletor2}, $I(t)$ stands for solar radiance focused upon the collectors (which is a load disturbance from a control viewpoint); $T_p$, $T_e$ and $T_f$ are, respectively, the collector plate, the external (load disturbance as well) and the fluid temperatures; $u$ is the inlet fluid flow, which is the control input of the system; finally, $A_{i}$ and $A_{e}$ are, respectively, the internal and external surfaces of the pipes, that have (internal and external) diameters of $d_i$ and $d_e$. 

For application purposes, as seen in \cite{pasamontes2013hybrid, ampuno2019apparent}, the space-derivative term $\frac{\partial T_{f}}{\partial s} (t,s)$ can be replaced by either a nonlinear function or an apparent transport delay. In this paper, it is approximated by the following nonlinearity
\begin{eqnarray}
\label{ApproxPartialDeriv}
\frac{\partial T_f(t,s)}{\partial s} & \approx &\frac{1-e^{-\frac{T_f (t)}{T_f^{max}}}}{(1-e^{-1})}  \, \text{,}
\end{eqnarray}
which means that the diffusion of the thermal energy of the fluid flowing along the flat collectors increases with respect to its temperature $T_f(t)$ until a certain level is attained $T_f^{max}$, after which the diffusion is constant, i.e. the whole fluid inside the flat collector is at the same temperature. This approximation is quite reasonable with respect to the ST application and in accordance with \cite{pasamontes2013hybrid}.

The heat transfer coefficient of the fluid $h_i\left(T_p(t)\right)$ is given according to the following nonlinear equation:
\begin{eqnarray}
\label{HiHeatTransfer}
h_i\left(T_p(t)\right) &=&\overline{h}_i\left(\frac{1-e^{\frac{-T_p(t)}{T_p^{max}}}}{1-e^{-1}}\right) \, \text{,}
\end{eqnarray}
where $\overline{h}_i$ is the maximal heat transfer coefficient of fluid, attained for $T_p(t) \, = \, T_p^{max}$, see \cite{pasamontes2013hybrid}.

\subsection{Model Parameters}

Regarding the nonlinear model of Eqs. \eqref{modfinalcoletor1}-\eqref{modfinalcoletor2} with the relaxations of Eqs. \eqref{ApproxPartialDeriv}-\eqref{HiHeatTransfer}, the parameters have been identified and adjusted for the CIESOL tested in previous papers \cite{morato2019SBAI}. The numerical values for these parameters are given in Table \ref{modelparametersandconstraints}. 

\begin{table}[htb]
    \caption{Model Parameters of the ST Process in Eqs. \eqref{modfinalcoletor1}-\eqref{modfinalcoletor2}.}
  \begin{center}
  \begin{tabular}{|c| c | c |c  |}
\hline
$\xi _m$ & $1100 \, \rm{kg/m^3}$ & $C _m$ & $440 \, \rm{J/(kg^oC)}$  \\
$\xi _f$ & $1000 \, \rm{kg/m^3}$ & $C_f$ & $4018 \, \rm{J/(kg^oC)}$
    \\ 
$A_{e}$ & $0.0038 \, \rm{m^2}$ & $A_{i}$ & $0.0013 \, \rm{m^2}$ \\
$d_i$ & $0.04 \, \rm{m}$ & $d _e$ & $0.07 \, \rm{m}$  \\
$h_0$ & $11$ & $\overline{h_i}$ & $800$ \\ 
$\nu $ & $3.655$ & $-$ & $-$ \\ \hline
  \end{tabular}
   \end{center}
\label{modelparametersandconstraints}\end{table}

\subsection{Performance Goals and Constraints}

The goal of this ST system is to track outlet temperature references to cover a certain heat demand, which is done by varying the inlet fluid flow $u$. This collector field has a $160 \, \rm{m^2}$ surface area, distributed in ten parallel rows composed of eight collectors per row. 

In terms of performances, the temperature set-point tracking should be done as fast as possible, while respecting the maximal temperature of $300\, \rm{^oC}$ that the inlet fluid can tolerate. Moreover, the temperature of the plates should not surpass $600\, \rm{^oC}$. These performances can be evaluated using usual reference-tracking indexes, such as the integral of the average tracking error.

The inlet flow (control effort) should be always positive (no fluid can be extracted from the ST units, only injected) and abide to a maximal value of $0.35 \, \rm{m^3/s}$, which is the upper constraint of the injection pump. Moreover, the control policy must be evaluated within $T_s \, = \, 3 \, \rm{s}$, which is the sampling period of this ST unit.

The disturbances to this system (the solar radiance and external temperature variables) are assumed to be measurable from a control viewpoint. This is quite reasonable, given that accurate estimations for the future behaviour of these disturbances can be indeed obtained \cite{camacho2012control}. These estimation results (for solar radiance and outside temperature) are easily provided with Neural Network tools, as seen in \cite{JoseJPC, rosiek2018online}.

Table \ref{ProcessConstraints} resumes the state and input constraints. Note that the fluid and plate temperatures are lower-bounded by external temperature to the ST system, $T_e(t)$. If there is no sun, the ST system will reach a thermal equilibrium with $T_e(t)$. For simplicity, since $T_e (t) > 0$, the lower bounds on $T_p$ and $T_f$ can be taken as $0$.

\begin{table}[htb]
    \caption{Constraints of the considered ST system.}
  \begin{center}
  \begin{tabular}{|c | c|}
\hline
$u(t) \, \in \, \mathcal{U}$ & $\mathcal{U} \, : = \, \left\{u \, \in \mathbb{R} \, | \, 0 \, \leq \, u \, \leq \, 0.35 \, \rm{m^3/s}\right\}$ \\ \hline
$T_p(t) \, \in \, \mathcal{T}_p$ & $\mathcal{T}_p \, : = \, \left\{T_p \, \in \mathbb{R} \, | \, T_e(t) \, \leq \, T_p \, \leq \, T_p^{max}\right\}$ , $T_p^{max} \, = \, 600 \, \rm{^oC}$ \\
$T_f(t) \, \in \, \mathcal{T}_f$ & $\mathcal{T}_f \, : = \, \left\{T_f \, \in \mathbb{R} \, | \, T_e(t) \, \leq \, T_f \, \leq \, T_f^{max}\right\}$ , $T_f^{max} \, = \, 300 \, \rm{^oC}$ \\\hline
  \end{tabular}
   \end{center}
\label{ProcessConstraints}\end{table}

\section{MPC Algorithms for Systems with LPV Models}
\label{sec2}

\subsection{LPV Model through Linear Differential Inclusion}

Throughout this paper, the focus is given to the control of a nonlinear process (the ST collector system), which can be embedded into an LPV setting. The LPV embedding of this nonlinear system is done through LDI.

For simplicity, consider the following discrete-time nonlinear model\footnote{The discrete-time iteration holds as $t = kT_s$, where $T_s$ is the sampling period of the system, $t \, \in \, \mathbb{R}^+$ is the time variable, while $k \, \in \, \mathbb{N}^+$ is the discrete indexing variable.}, which is a discretized version of Eqs. \eqref{modfinalcoletor1}-\eqref{modfinalcoletor2} with the relaxations of Eqs. \eqref{ApproxPartialDeriv}-\eqref{HiHeatTransfer}, where $x \in \mathbb{R}^x$ are measurable states, $u \in \mathbb{R}^u$ is the control signal and $w \in \mathbb{R}^w$ are load disturbances:
\begin{eqnarray}
\label{LDINON}
x(k+1) &=& f_x(x(k),u(k),w(k)) \,\text{.} 
\end{eqnarray}

In fact, since the considered ST system operates under a $T_s \, = \, 3 \, \rm{s}$ sampled period, Eq. \eqref{LDINON} is found through Euler discretization of the real nonlinear model (with the discussed relaxations). 

Then, LDI is verified if for each $x$, $u$ and $w$ and every instant $k$, there exists a matrix $G(x,u,w,k) \in \mathcal{G}$ such that:
\begin{eqnarray}
\label{LDINONLPV}
\left[\begin{array}{c} f_x(x(k),u(k),w(k)) \end{array}\right]&=& G(x,u,w,k)
                                           \left[\begin{array}{c} x(k)
                                                   \\ u(k) \\ w(k) \end{array}\right]\,\text{,}
\end{eqnarray}
where $\mathcal{G} \in \mathbb{R}^{(x)\times(x+u+w)}$ is the LDI matrix.

Indeed, the LDI property holds for the ST nonlinear discrete-time system in Eq. (\ref{LDINON}), considering the system states as $x(k) \, = \, \left[\begin{array}{cc} x_1(k) & x_2(k) \end{array}\right]^T \, = \, \left[\begin{array}{cc} T_p(k) & T_f (k) \end{array}\right]^T$. This means that the system can be represented within an LPV framework. Under such LPV formalism\footnote{Note that, although this paper proposes a formulation for LPV systems, this model could also be used for the case of linear time-varying (LTV) systems.}, the following model is found:
\begin{eqnarray}\label{eq:LPVsystem}
    x(k+1)& =& A(\rho(k)) x(k) + B(\rho(k)) u(k) + B_w w(k)\, \text{,}\\
    \rho (k) &=& f_\rho(x(k)) \, \text{,}
\end{eqnarray}
where $f_\rho$ represents the endogenous nonlinear function for the evolution of the scheduling parameters, $[A(\rho),B(\rho)]$ are the matrices of this model, affine on the scheduling term $\rho$. The vector of load disturbances $w(k)$ stands for $\left[\begin{array}{cc} I(k) & T_e(k)\end{array}\right]^T$. 

The LPV scheduling parameter $\rho = [\rho_1 , \rho_2]^T$ derives directly from the nonlinearities added to the balance of energy equations due to the time-varying thermal loss map of Eq. \eqref{HiHeatTransfer} and the partial derivative approximation of Eq. \eqref{ApproxPartialDeriv}. Thus, the endogenous nonlinear map for the scheduling parameters is the following:
\begin{eqnarray}
\left[\begin{array}{c} \rho_1(k) \\ \rho_2(k)  \end{array}\right]^T \, = \, f_\rho(x(k)) &=& \left[\begin{array}{c} d_i\pi\overline{h_i}\left(\frac{1-e^{-\frac{x_1(t)}{T_p^{max}}}}{1-e^{-1}}\right) \\ \frac{1-e^{-\frac{x_2(t)}{T_f^{max}}}}{(1-e^{-1})A_i} \end{array}\right] \, \text{.}
\end{eqnarray}

Consequently, each of the scheduling parameters is bounded to a convex set:
\begin{eqnarray}
\rho_1 \, \in \, [\underline{\rho_1} \, , \, \overline{\rho_1}] &= & [0 \, , \, d_i\pi\overline{h_i}] \, \text{and} \\ 
\, \rho_2 \, \in \, [\underline{\rho_2} \, , \, \overline{\rho_2}]  &=& \left[0 \, , \, \frac{1}{A_i}\right] \, \text{,}
\end{eqnarray}
which means that $\rho \, \in \, \mathcal{P}$.

The matrices of the LPV model in Eq. \eqref{eq:LPVsystem} are analytically given\footnote{Notation $\mathbb{I}_x$ stands for the identity matrix of dimension $x$.}:
\begin{eqnarray}
\label{matrizesmodeloqLPV}
A(\rho) &=& \mathbb{I}_{x}+ T_s\left[\begin{array}{cc} - \frac{d_e\pi h_0}{\epsilon _m C_m A_e} - \frac{1}{\epsilon _m C_m A_e}\rho _1 & \frac{1}{\epsilon _m C_m A_e}\rho _1 \\ \frac{1}{\epsilon _f C_f A_i}\rho _1 & -\frac{1}{\epsilon _f C_f A_i}\rho _1 \end{array} \right] \, \\ 
B(\rho) &=& T_s \left[\begin{array}{c} 0 \\ -\rho_2 \end{array}\right] \, \text{,} \\
B_w &=& T_s\left[\begin{array}{cc} \frac{d_e\pi \nu}{\epsilon _m C_m A_e} & \frac{d_e\pi h_0}{\epsilon _m C_m A_e} \\ 0 & 0\end{array} \right] \, \text{.}
\end{eqnarray}

Regarding this LPV formulation for the nonlinear ST plant process, and due to the hard physical constraints given in Table \ref{ProcessConstraints}, it follows that the LPV model should be regulated with respect to hard constraints on the state and control vectors, due to operation feasibility of the system, as given:
\begin{eqnarray}
\label{eq:HardConstraints}
x(k) \in \mathcal{X} \, = \, \mathcal{T}_p \times \mathcal{T}_f \quad \text{and}  \quad u(k) \in \mathcal{U} \quad \text{for all } k \geq 0 \, \text{,}
\end{eqnarray}
which are convex and compact subsets of $\mathbb{R}^x$ and $\mathbb{R}^u$, respectively. Both of these sets contain the origin, i.e. they are proper $\mathcal{C}$ sets.

In addition, throughout the sequel of this paper, it follows that for all $k$:
\begin{eqnarray}\label{eq:Polytope}
        [A(\rho(k)), B(\rho(k))] &\in& \Omega \, \text{,}
\end{eqnarray}
where $\Omega$ is a polytope that represents Eq. \eqref{eq:LPVsystem} as LTI models at its $L = 4$ vertices, which can be represented as:
\begin{eqnarray}
    \Omega &=& \text{Co}\{[A_1, B_1],[A_2, B_2], \dots , [A_L, B_L]\} \, \text{,}
\end{eqnarray}
where Co$\{\cdot \}$ denotes a convex hull and $[A_j, B_j]$ are the LTI model matrices of the hull. 

Note that the LTI model matrices are trivially found for the four possible combinations of the two scheduling parameters $\rho_1$ and $\rho_2$ (at their minimal and maximal values), as follows:
\begin{eqnarray}
\label{A1B1matrix}
A_1 \, = \, A(\rho)|_{\rho = \left\{ \underline{\rho_1} \, , \, \underline{\rho_2}\right\}} \quad \text{and} \quad B_1 \, = \, B(\rho)|_{\rho = \left\{ \underline{\rho_1} \, , \, \underline{\rho_2}\right\}} \, \text{,}\\
A_2 \, = \, A(\rho)|_{\rho = \left\{ \underline{\rho_1} \, , \, \overline{\rho_2}\right\}} \quad \text{and} \quad B_2 \, = \, B(\rho)|_{\rho = \left\{ \underline{\rho_1} \, , \, \overline{\rho_2}\right\}} \, \text{,}\\
A_3 \, = \, A(\rho)|_{\rho = \left\{ \overline{\rho_1} \, , \, \underline{\rho_2}\right\}} \quad \text{and} \quad B_3 \, = \, B(\rho)|_{\rho = \left\{ \overline{\rho_1} \, , \, \underline{\rho_2}\right\}} \, \text{,}\\ \label{A4B4matrix}
A_4 \, = \, A(\rho)|_{\rho = \left\{ \overline{\rho_1} \, , \, \overline{\rho_2}\right\}} \quad \text{and} \quad B_4 \, = \, B(\rho)|_{\rho = \left\{ \overline{\rho_1} \, , \, \overline{\rho_2}\right\}} \, \text{.}
\end{eqnarray}

\begin{figure}[ht]
\centering
\includegraphics[width=0.5\linewidth]{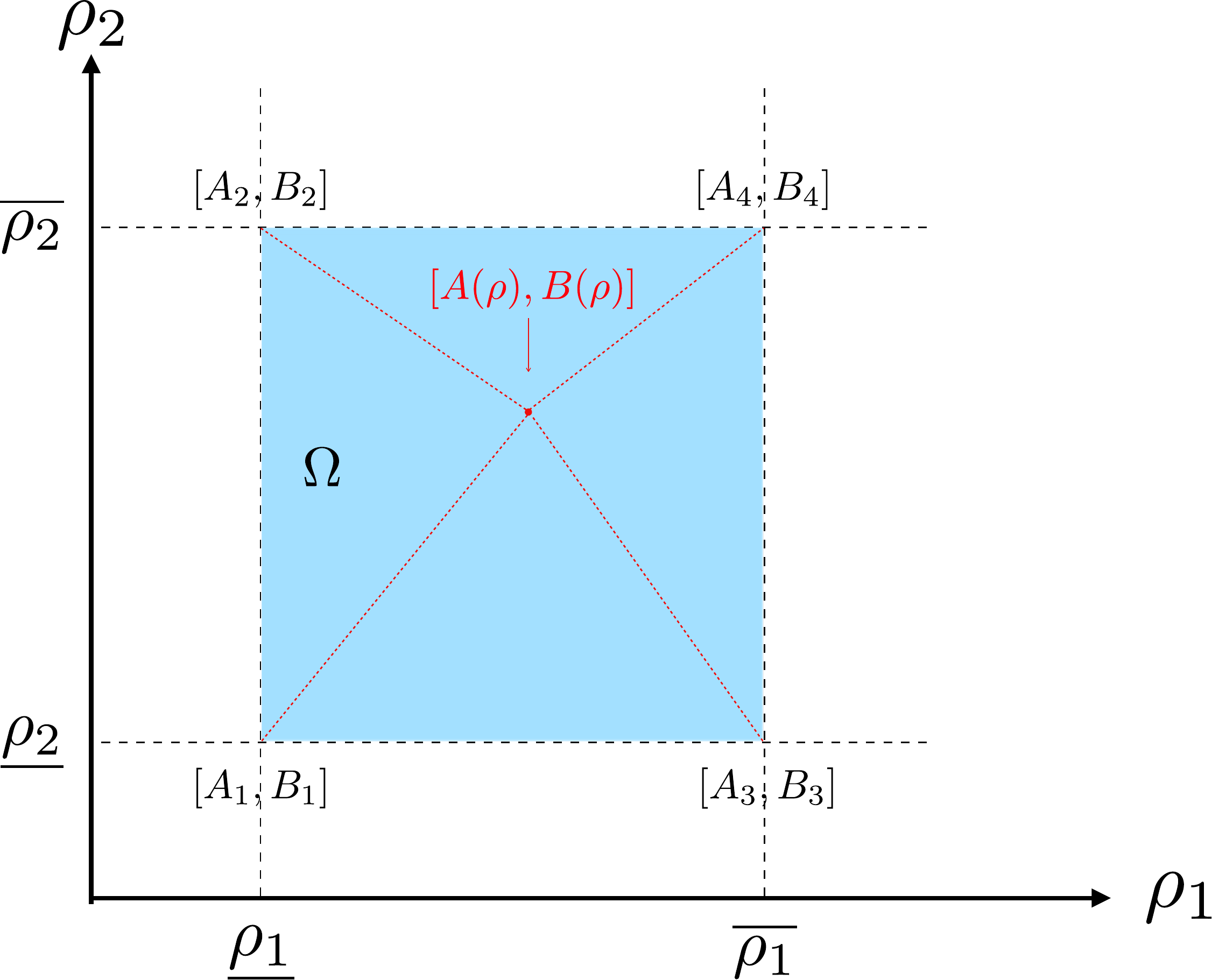}
\caption{\label{thePolytope} Polytopic Representation of an LPV Model with two scheduling parameters; $\Omega$ is the polytope.}
\end{figure}

Figure \ref{thePolytope} illustrates how the convex polytope $\Omega$, defined as the hull of the four matrix pairs $[A_j,B_j]$ is found. Each vertex is an LTI model, whereas the LPV model is a combination of these four models.

\subsection{Predictive Control through an LPV Model}

The complete standard model-based predictive control algorithm is capable of obtaining an optimal control law that takes into account constraints on the states, outputs and control actions. Widely used for performance regulation, this control procedure resides in solving the following problem\footnote{Notation $(k+i|k)$ represents a prediction for instant $k+i$, computed at instant $k$.}:
\begin{problem}
\label{PrbMPC}
\begin{eqnarray}
\label{PrbMPCeq}
\min_{\mathbf{u}} \, V_N &=& \min_{\mathbf{u}} \,\sum_{i=1}^{N}\ell
                         \left(x(k+i|k),u(k+i-1|k)\right) \\
\label{constSModel} \text{s.t.} && \text{System Model} \, \text{,}\\ 
&& \label{ConstraintU} u(k+i-1|k) \, \in \, \mathcal{U} \quad \forall i \, \in \, \mathbb{Z}_{1:N} \, \text{,}\\
&& \label{ConstraintX} x(k+i|k) \, \in \, \mathcal{X} \quad \forall i \, \in \, \mathbb{Z}_{1:N} \,  \text{,}
\end{eqnarray}
where $\mathbf{u}$ is the optimal sequence of control actions along the prediction horizon. 
The stage cost $\ell(\cdot)$ weights the control action and the states at each future instant; this function is usually quadratic on $x$ and $u$. Additionally, a terminal 
cost and output constraints are sometimes considered, as well as the use of terminal 
and slew rate constraints, on $\delta u (k+i|k) =  u(k+i|k) - u(k+i-1|k)$.
\end{problem}

When the MPC Problem \ref{PrbMPC} is applied to a nonlinear process with an LPV model, the evolution of the scheduling variables along the prediction horizon $N$ becomes necessary to describe the future values of the states. Since the nonlinear map $f_\rho (\cdot)$ gives the evolution of the endogenous scheduling variables, from the viewpoint of instant $k$, the next $N$ iterations of this map are:
\begin{eqnarray}
\label{SchedulingFuture}
\Gamma _k &=& \text{col}\{\rho(k+1) \, , \, \rho(k+2) \, , \, \dots \, , \,
           \rho(k+N-1)\} \, \text{.}
\end{eqnarray}

Then, for any admissible initial condition $x(k) = x_k \, \in \, \mathcal{X}$, the standard MPC procedure, given by Problem \ref{PrbMPC}, for the considered LPV embedding of the ST system in Eq. \eqref{eq:LPVsystem}, has to internally elaborate the model prediction constraint from Eq. \eqref{constSModel}, which exhibits nonlinearities from the second iteration onward (neglecting $w$):
\begin{eqnarray}
\label{pred2}
x(k+2|k) = A(\rho(k+1))A(\rho(k))x_k + A(\rho(k+1))B(\rho(k))u(k|k) + B(\rho(k+1))u(k+1|k).
\end{eqnarray}
and so forth, up to the $N$-th iteration. This results, therefore, in an NP version of Problem \ref{PrbMPC}. 

Considering the goal of applying MPC to the nonlinear ST process, the LPV framework offers some advantages that can be used to simplify this NP into a QP formulation. In a regular NMPC formulation, it would be imperious to know the exact behaviour of the process model $f_x(\cdot)$ along the prediction horizon. In the pure LPV embedding case, this converts into the necessity of the future values of the scheduling parameter along $N$, coupled as $\Gamma_k$. The advantage of the LPV setting appears with regard to $\Gamma_k$, since the LPV model can be described, for all future instants $k+n$, by a generic pair $[A(\rho(k+n)), B(\rho(k+n))]$ which belongs to the polytope $\Omega$. Therefore, any pair $[A(\rho(k+n)), B(\rho(k+n))]$ can be represented as a convex combination of the LTI vertices of this polytope as follows:
\begin{eqnarray}
\label{eq:NewConstraints}
        A(\rho(k+n)) \,=\, \displaystyle\sum_{j=1}^{L} \mu_j(k+n) A_j \, \text{ and } \, B(\rho(k+n)) \,=\, \displaystyle\sum_{j=1}^{L} \mu_j(k+n) B_j \, \text{,}\\ \label{eq:NewConstraints2}
        \text{with } \displaystyle\sum_{j=1}^{L} \mu_j(k+n) \,=\, 1 \, \text{ and } 0 \leq  \mu_j(k+n) \,\leq\, 1 \, \text{,} \quad j \in \mathbb{Z}_{1:4} \, \text{,} 
\end{eqnarray}
with, for $L=4$, since the system has two scheduling parameters:
\begin{eqnarray}
\label{describemus}
\mu_1 (k+n) &=& \left(\frac{\overline{\rho_1} - \rho_1(k+n)}{\overline{\rho_1} - \underline{\rho_1}}\right) \left(\frac{\overline{\rho_2} - \rho_2(k+n)}{\overline{\rho_2} - \underline{\rho_2}}\right) \, \text{,} \\ \nonumber
\mu_2 (k+n) &=& \left(\frac{\overline{\rho_1} - \rho_1(k+n)}{\overline{\rho_1} - \underline{\rho_1}}\right) \left(\frac{\rho_2(k+n) - \underline{\rho_2}}{\overline{\rho_2} - \underline{\rho_2}}\right) \, \text{,} \\ \nonumber
\mu_3 (k+n) &=& \left(\frac{\rho_1(k+n) - \underline{\rho_1}}{\overline{\rho_1} - \underline{\rho_1}}\right) \left(\frac{\overline{\rho_2} - \rho_2(k+n)}{\overline{\rho_2} - \underline{\rho_2}}\right) \, \text{,}\\ \nonumber
\mu_4 (k+n) &=& \left(\frac{\rho_1(k+n) - \underline{\rho_1}}{\overline{\rho_1} - \underline{\rho_1}}\right) \left(\frac{\rho_2(k+n) - \underline{\rho_2}}{\overline{\rho_2} - \underline{\rho_2}}\right) \, \text{.} 
\end{eqnarray}

Note that each $\mu_j(k+n)$ is a weighting variable that determines how much does the $j$-th vertex of the scheduling polytope (LTI model) represents the LPV model at a given future instant $k+n$. 

Therefore, due to this polytopic characterist of the LPV embedding of the ST process, $\Gamma_k$ can be replaced in the MPC Problem \ref{PrbMPC} by the respective convex sum of these four LTI models, which are always known. If the four weighting variables $\mu_j(k+n)$ are assumed to be known along $N$, the NP is converted into a QP version, which can be evaluated much faster than full-blown NMPC procedures.

For notation simplicity, $\mu$ denotes hereafter the vector that collects these four weighting variables, i.e. $\mu(k) \, = \, \text{col}\{\mu_1(k) \, , \, \mu_2(k)\, , \, \mu_3(k) \, , \, \mu_4(k) \}$.

\begin{remark}
Instead of using the polytopic representation of the LPV embedding, one could also admit that $\rho$ is known for all future instants inside the $N$ horizon. By doing so, the model-based predictions in Eq. \eqref{pred2} would be converted into a linear formulation. To say one has knowledge of the complete future scheduling vector $\Gamma _k$ is obviously false, since only the instantaneous value of $\rho$, i.e. $f_\rho(x(k))$ is known. Thus, in \cite{morato2018lpv,cisneros2018constrained, MoratoLPVMPCIJEPES, Morato2020qLPVMPC}, different estimation strategies are used to provide a frozen guess $\hat{\Gamma}_k$ at each instant $k$ to substitute $\Gamma_k$ in the MPC Problem and render it as QP version of this control problem.
\end{remark}

\section{Adaptive LPV MPC Method}
\label{sec3}

With the previous formalities in mind, this Section presents a novel adaptive MPC design procedure for a ST unit modelled under an LPV formalism. This AMPC regulation policy tries to find, at each sampling instant $k$, the best LTI prediction model for the next $N$ steps, based on the previous $N$ steps of data, and uses some terminal ingredients to guarantee stability. The basic idea of behind this novel method is to consider the LPV polytopic combination variables $\mu_j$ from Eqs. \eqref{eq:NewConstraints}, \eqref{eq:NewConstraints2} and \eqref{describemus} as virtual weights that, at each sampling instant, indicate which is the best LTI combination model that can be used to momentarily describe the controlled ST process (based on the previous measurement and the desired set-point).

Therefore, from now on, the vector $\mu$ is treated as a new decision variable of the optimization problem and Eqs. \eqref{eq:NewConstraints}-\eqref{eq:NewConstraints2} become constraints of the procedure, together with the feasibility regions given by the hard constraints in Eq. \eqref{eq:HardConstraints}.

In this sense, this novel LPV MPC method adapts the process model to the uncertain system into a single LTI prediction model, at each sampling instant, weighting the LTI vertex of polytope $\Omega$, those given through Eqs. \eqref{A1B1matrix}-\eqref{A4B4matrix}, through $\mu$ to find the ``ideal'' one for the prediction of state variables for the following $N$ steps. This prediction model is based on the data set of the a backward horizon comprising the previous $N$ steps. Synthetically, the procedures tries to find the best LTI model to match the backward horizon dataset and then uses this model to make the predictions for the forward horizon, at each sampling instant.

Due to the simplification of finding an LTI prediction model, the major consequence is of having model-process mismatches regarding the control horizon. This means that the proposed method is obviously \textit{a priori} sub-optimal, which does not mean that the achieved results will not be very near the optimality condition. Moreover, a great advantage of using a simplified prediction model is that this predictive control frameworks yields one QP for ``identification'' purposes (regarding $\mu$) and another QP for control purposes, which can be solved online with fast solvers and used for a real-time implementation of this control strategy for ST collectors.

\subsection{Backward MHE QP}

The backward QP is used to find a constant vector $\mu$ that best matches the polytopic LPV model to the real data. This procedure is indeed very simple: it minimizes the model-data discrepancy with respect to $\mu$ and the variance of $\mu$ along the simulation.

The virtual tuning variable is found with the solution of the following optimization problem, from $k = k_0$, considering $x$ and $u$ as measured data and $\mu_{k-1}$ as the result from the previous iteration:
\begin{eqnarray} \label{BackwordOptimizationProblemMHE}
    \min_{\mu}&& \sum _{i = k-N +1}^{k} \left( e(i)^T Q_e e(i) + \nu_{\mu}^T Q_{\nu} \nu_{\mu} \right)\\ \nonumber
    \textbf{s.t.} & &\, \\
    \underbrace{e(i+1)}_{\text{Model-matching Error}} &=& \overbrace{x(i+1)}^{\text{Known data}} - \overbrace{\left(A(\rho) x(i) + B(\rho) u(i) + B_w w(i)\right)}^{\text{LPV Model}} \quad i \in \mathbb{Z}_{k-N:k-1}\, \text{,} \\
\label{newconst1}
    A(\rho) &=& \displaystyle\sum_{j=1}^{4} \mu_j A_j ~~ \text{and}  ~~ B(\rho) \,\,=\,\, \displaystyle\sum_{j=1}^{4} \mu_j B_j~~\quad j \in \mathbb{Z}_{1:4}\, \text{,} \\ 
    \displaystyle\sum_{j=1}^{4} \mu_j &=& 1 \quad j \in \mathbb{Z}_{1:4} \, \text{,}\\ \label{newconst4}
    0 &\leq&  \mu_j \leq 1 ~~\qquad j \in \mathbb{Z}_{1:4} \, \text{,} \\
    \mu &=& \text{col}\{\mu_j(k) \} ~~\quad j \in \mathbb{Z}_{1:4}  \, \text{,} \label{newconst3} \\ 
    \mu &=& \mu_{k-1} + \nu_{\mu} ~~ \, \text{.}
\end{eqnarray}

Matrices $Q_e$ and $Q_\nu$ are tuning weights of this optimization procedure. For simplicity, they are taken as identity matrices.

Indeed, note that the above QP works exactly as the MHE scheme proposed in the literature \cite{rawlings2006particle, kuhl2011realparameter} for the estimation of time-varying parameters.

\subsection{Forward MPC QP}

Considering the regulation, the forward MPC procedure is a standard QP that slightly differs from Problem \ref{PrbMPC}. In this paper, the ``MPC for Tracking'' method \cite{limon2008mpc} is considered for regulation purposes: this control design is used to ensure that the controller can asymptotically lead the process to a steady-state reference $x_s$ in an admissible trajectory from any feasible initial state $x_0$. The approach consists basically in adapting the standard MPC
cost function (i.e. weighting the quadratic difference between 
output and reference). 

\begin{remark}
The ``MPC for Tracking'' design embeds an artificial reference $x_a$ to the optimization problem and sets the system states to track this artificial variable. Altogether, it determines that this artificial set-point should be as close as possible to the real state reference $x_s$, which altogether ensures an enlarged domain of attraction. The target operation point $p_s \, = \, (x_s,u_s)$ is an admissible steady-state for Eq. \eqref{eq:LPVsystem}. 
\end{remark}

\begin{assumption} 
\label{ChoiceofP}
Consider: (1) $Q \, \in \, \mathbb{R}^{x \times x}$ and $R \, \in \, \mathbb{R}^{u \times u}$ as positive definite matrices; and (2) $\kappa \, \in \, \mathbb{R}^{u \times x}$ as an arbitrary stabilizing state-feedback control gain of the process model. For these matrices, it is implied that, for the ST  discrete-time LPV model, $A(\rho(k))+B(\rho(k))\kappa$ is Schur. Then, there exists another positive definite matrix  $P \, \in \, \mathbb{R}^{x \times x}$ such that $$(A(\rho_k)+B(\rho_k)\kappa)^T P  (A(\rho_k)+B(\rho_k)\kappa) -P = - (Q + \kappa^TR\kappa)$$ holds for all $\rho_k \, \in \, \mathcal{P}$ (note that the disturbance vector $w(k)$ is neglected).
\end{assumption}

Then, as long as the previous Assumption hold, the MPC problem is formulated with the following cost function, considering that $\mathbf{\mu}$ represents the value obtained with the backward MHE QP for $\mu$, i.e.:
\begin{eqnarray}
\label{eq:CostFunction}
    V_N(x, \mathbf{\mu};\mathbf{u}) &=& \|x_a - x_s \|_{T_x}^2 +  \|x(N) - x_a \|_P^2 + \sum_{k=1}^{N-1} \left(\|x(k) - x_a \|_Q^2\right) + \sum_{k=0}^{N-1}\left(\|u(k) - u_s \|_R^2\right) 
    \, \text{.}
\end{eqnarray}

In this stage cost $V_{N}(\cdot)$, $x_a \, \in \, \mathcal{X}$, $u_s \, \in \, \mathcal{U}$ define an artificial target regulation goal $p_a$ and the term $\|x(N) - x_a \|_P^2$ is an offset that penalizes the final-state deviation from this target operation $p_a$. Moreover, the offset term $\|x_a - x_s \|_{T_x}^2$ ensures that the artificial variable tracks the real set-point variable, with the actual target goal defined by $p_s$. Note that the inclusion of this suitable penalization of the terminal state $x(N)$ can steer to asymptotic stability with good performances, as evidenced in \cite{ferramosca2009mpc}.

\begin{remark}
The objective of the inclusion of the artificial target point $p_a$ works as follows. Consider that the system evolves as predicted (with $\mathbf{\mu}$ representing the weight for the LTI models) and that the actual target point $p_s \, = \, (x_s,u_s)$ is an admissible point contained inside the
tracking set $\mathcal{T} \, : = \, \mathcal{X} \times
\mathcal{U}$ and that it can be tracked within $N$ steps. Then, $p_a$ becomes an asymptotically stable point in
closed-loop, since the MPC will ensure convergence to it. If the system cannot ensure that the target reference $p_s$ is tracked within the horizon of $N$ steps, then the artificial reference $x_a$ enables it to stabilize at more options of closed-loop equilibria points, as close as possible to $x_s$, since $x_a$ is set to converge to the actual target. If $x_s$ is not trackable, then the achieved closed-loop equilibrium is given by $p_s^\star \, = \, (x_s^\star,u_s^\star) \, = \, \text{arg
  min}_{x_a}\|x_a - x_s \|_{T_x}^2$. Note that the weighting matrix $T_x$ is taken as a parametrized version of $P$, i.e. $T_x \, = \, \alpha _{T_x} P$. In this paper, for simplicity, $\alpha_{T_x}$ is taken as an identity block.
\end{remark}




Finally, the controller derived with this adaptive method is found with the solution of the following optimization problem, from $k = k_0$:
\begin{eqnarray} \label{eq:OptimizationProblem}
    \min_{\mathbf{u}}&& V_N(x,\mathbf{\mu};\mathbf{u}) \\ \nonumber
    \textbf{s.t.} \, \, \,  x(0)&=&x(k_0) \, \text{,}\\
    x(k+1) &=& A x(k) + B u(k) + B_w w(k) \quad k \in \mathbb{Z}_{0:N-1}\, \text{,} \\
\label{newconst11}
    A &=& \displaystyle\sum_{j=1}^{4} \mu_j A_j ~~ \text{and}~~ B \,\,= \, \,\displaystyle\sum_{j=1}^{4} \mu_j B_j ~~\quad j \in \mathbb{Z}_{1:4} \, \text{,}\\ 
    x(k) &\in& \mathcal{X}, \quad \forall k \in \mathbb{Z}_{0:N-1} \, \text{,}\\
    u(k) &\in& \mathcal{U}, \quad \forall k \in \mathbb{Z}_{0:N-1} \, \text{,}\\
    x(N) &\in& \mathbf{X}_f \, \text{,}
\end{eqnarray}
where $\mathbf{X}_f$ is an adequate robust controlled positively terminal invariant set that contains $p_s$. To determine this invariant set, some suggestions are also given in \cite{morato2019LPVS} and \cite{pipino2019mpc}. Synthetically the essential idea on how to determine this robust invariant set for the LPV system in Eq.  \eqref{eq:LPVsystem} is to find $\mathbf{X}_f \, \subset \, \mathcal{X}$ so that for all
possible $x(k_0) \in \mathbf{X}_f$ there must exist a feasible input $u =
\kappa(x(k_0)) \in \mathcal{U}$ which guarantees that $x(k_0+1)$ lies inside $\mathbf{X}_f$ despite model-process mismatches. Since the LPV system is polytopic inside $\Omega$, this set is computed as the intersection of the robust invariant sets for the four LTI vertices through the control matrices $[A_j \, , \, B_j]$.

By considering a receding horizon policy, the proposed regulation control policy that is obtained by solving the above optimization is given by: 
\begin{align*} 
    \kappa(x(k_0)) &= \mathbf{u}^\star(0;x(k_0)) \, \text{,}
\end{align*}
being $\mathbf{u}^\star$ the solution of the second QP, which represents the optimal sequence of control actions to be applied for reference tracking purposes.

Note that in the optimization procedure of Eq. \eqref{eq:OptimizationProblem}, the future values for $w(k)$ are necessary. As previously discussed, the instantaneous and future values for solar radiance and external temperature can be found through estimation algorithms \cite{JoseJPC, rosiek2018online}.

\section{Robust Tube-based Method}
\label{sec4}

One of the motivations of this paper is to compare distinct methods of predictive control works that have been applied for ST collector systems. For this goal, the proposed (sub-optimal) AMPC technique will be compared to a robust MPC (RMPC) policy, which is based on a set of trajectory tubes to embed the nonlinearities. Therefore, this Section rapidly recalls the tube-based MPC (TMPC) design procedure, from the literature.

Generally speaking, robust control policies are those able to steer the system to a specified target despite uncertainties. RMPC methods \cite{Mayne:4} are based on worst-case optimization procedures, taking into account the whole uncertainty set. RMPC fits to the purpose of nonlinear and LPV processes, since these can roughly be represented as LTI ones with known associated uncertainties; take the following polytopic LPV model, as the one of the considered ST unit:
\begin{eqnarray}\label{modelqLPV}
x(k+1) &=& A(\rho)x(k) + B(\rho)u(k) + B_w w(k)\, \text{.}
\end{eqnarray}

Writing $A(\rho) = A_0 + A_1(\rho-\overline{\rho})$ and $B(\rho) = B_0 + B_1(\rho-\overline{\rho})$, being $A_0 = A(\overline{\rho})$ and $B_0 = B(\overline{\rho})$ with $\overline{\rho}$ as an arbitrary (average) frozen value for $\rho$, it yields:
\begin{eqnarray}\label{modelLTIunc}
x(k+1) &=& A_0x(k) + B_0u(k) + B_w w(k) + \xi(k) \, \text{,}
\end{eqnarray}
with $\xi (k) = A_1(\rho-\overline{\rho})x(k) + B_1(\rho-\overline{\rho})u(k) \in \mathcal{E}$ as the bounded uncertainties. As demonstrated in \cite{gesser2018robust}, Eq. \eqref{modelLTIunc} is exactly the model used for RMPC design.

RMPC is rather consolidated; literature shows a range of works that vary according to how the optimization problem is set up and how the uncertainty set $\mathcal{E}$ is described \cite{Mayne:4,vesely2010robust,mayne2005robust}. Anyhow, as previously discussed, sometimes it is not clear for the designer if it is best to apply an RMPC tool or a simpler near-optimal algorithm when considering the case of ST processes. TMPC is a good example of how RMPC works and can serve for the comparison with the proposed technique.

This control framework, as proposed by \cite{Mayne:2}, resides on the fact that both open-loop and closed-loop versions of processes subject to unknown (but bounded) uncertainties generate a finite set of possible trajectories. These trajectories are often called "tubes" and correspond to a specific realization of the uncertainty set. In the tube-based design paradigm, the controller must compute a tube such that all state possible trajectories remain inside this feature for all possible realizations of the (bounded) disturbances and uncertainties, while guaranteeing the satisfaction of all control specifications.

To apply this method, the LPV model in Eq. \eqref{eq:LPVsystem} must be adapted (for now, the disturbances are neglected, since they are linearly included to the model). Since this polytopic model matrices are affine on the scheduling parameter $\rho$, the system can be re-written on the form of Eq. \eqref{modelLTIunc}; the parametric model uncertainty $\xi (k)$, if null, translates this LPV model into an  LTI one. Moreover, since $\rho$, $x$ and $u$ are ultimately bounded, it holds that $\xi \, \in \, \mathcal{E}$ for all $k \geq 0$.

For the tube-based design procedure, it is considered that there exists a corresponding nominal system to Eq. \eqref{modelLTIunc}, which is defined as:
\begin{eqnarray}
\label{modelz}
z(k+1) &=& A_z z(k) + B_z v(k)  + B_w w(k)\, \text{,}
\end{eqnarray}
where $z(k) \in \mathcal{Z}$ is the nominal state and  $v(k) \in \mathcal{V}$ is the control policy for the nominal system. The technique implies that, if the deviation between the real state and the nominal $e(k) = x(k) - z(k)$ is inside a robust invariant set $\mathbf{X}_i$, then the actual state is inside a tube enclosure and is, therefore, controllable.

TMPC, essentially, traces the trajectories for the LPV system as if it was an LTI one with bounded uncertainties $\xi (k)$. Then, if the actual trajectories fall under this envelope, the system can be steered to the desired equilibrium $p_s$. The state-feedback policy for this method is given by $u(k) = v(k) + K_te(k)$. If $x(k) \in \mathcal{X}$, $u(k) \in \mathcal{U}$ and $\xi (k) \in \mathcal{E}$ are satisfied when this control law is used, the tighter constraints for the nominal system are also satisfied \cite{gesser2018robust}:
\begin{eqnarray}\label{setZ_V}
	z(k) \in \mathcal{Z} &=& \mathcal{X} \ominus \mathbf{X}_i \, \text{,} \\
	v(k) \in \mathcal{V} &=& \mathcal{U} \ominus K_t\mathbf{X}_i \, \text{,}
\end{eqnarray}
where $\ominus$ is the Minkowski difference operator. 

For a TMPC method to work properly, a suitable nominal trajectory is required beforehand. Thus, an optimization procedure is used to obtain this adequate nominal trajectory envelope, considering the nominal system:
\begin{equation*}
\min_{\{v(j)\}}\sum_{j=0}^{N-1} \ell\left(z(j+1),v(j)\right)
\end{equation*}
\begin{equation}\label{optimization3}
\begin{aligned}
\text{s.t.}   \hspace{0.5cm} & \text{Model Eq. \eqref{modelz}}, && \\
\hspace{0.5cm} &z(j)\in \mathcal{Z},   &&j > k\\
&v(j)\in \mathcal{V},   &&j \geq k  \\
&z(N)\in \mathbf{Z}_f  &&
\end{aligned}
\end{equation}
For further details, refer to \cite{mayne2011tube}.



\section{Simulation Results}
\label{sec5}

Since the proposed AMPC method has been thoroughly explained and a TMPC has also been recalled, this Section presents realistic simulation results of the considered ST system. These two control methodologies are compared against each other but also to a simpler linearization-based MPC (denoted LTIMPC), which solves the QP in Eq. \eqref{eq:OptimizationProblem} disregarding the terminal set constraint and taking all vertice weighting variables as $\mu_j= 0.25$, which is an average linear model for the ST system, considering the four vertices of the polytope $\Omega$. This third technique can be understood as a nominal, averaged LTI MPC paradigm, since it takes the tuning variables $\mu_j$ as equally weighted. As might be expected, this controller is not able to achieve successful results for a large operation region, and it might even lead to infeasibility, since it takes the four model-tuning variables as $\mu_j = 0.25$, which obviously does not represent the whole scheduling polytope $\Omega$.

\begin{figure}[ht]
\centering
\includegraphics[width=\linewidth]{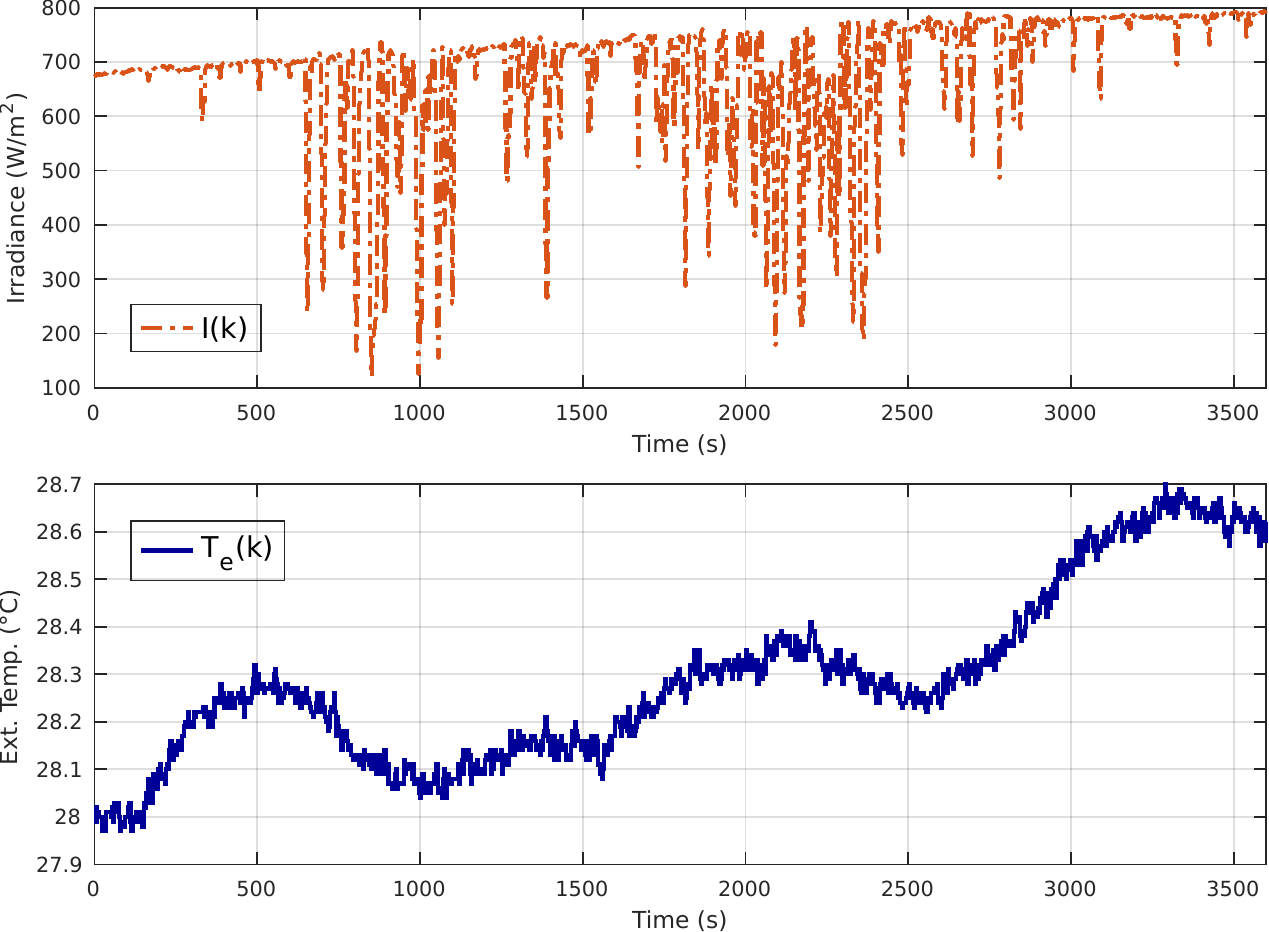}
\caption{\label{DisturbsFig} Considered Load Disturbances $w(k)$: Solar Irradiance and External Temperature.}
\end{figure}

The following results comprise the constrained regulation of the outlet temperature $x_2$, despite variations upon the solar radiance and outside temperature disturbances $w$. Real meterological data from the region of the CIESOL testebed is used for the solar radiance and temperature disturbances, considering a full hour of simulation \cite{hernandez2017use}. These disturbances are known form a control viewpoint; they are given in Figure \ref{DisturbsFig}. 

The ST system is emulated using a realistic, high-fidelity model, considering the nonlinear Eqs. \eqref{modfinalcoletor1}-\eqref{modfinalcoletor2}. Recall that these nonlinear equations have been previously validated to thoroughly emulate the CIESOL ST system \cite{pasamontes2013hybrid}. The controllers were synthesized using Matlab with Yalmip toolbox and \textit{QuadProg} QP solvers. 

The reference tracking goal is set as $97$ C for the fluid temperature $x_2(k)$, and $109.93$ C for the plate temperature $x_1(k)$. In fact, the hard-constrained set-point (SP) is the one for $x_2$, which must be tracked with minimal error as possible. Notice that integral action is not necessary in the MPC Problem \ref{PrbMPC}, since $\pm 0.5$ C is tolerated. If sought, integral action could be easily included by defining a new constraint $u(k+i) = u(k+i-1) + \delta u (k+i)$. Note that all three methods guarantee that the constraints on $x$ and $u$ (given in Table \ref{ProcessConstraints}) are respected.

\begin{figure}[ht]
\centering
\includegraphics[width=\linewidth]{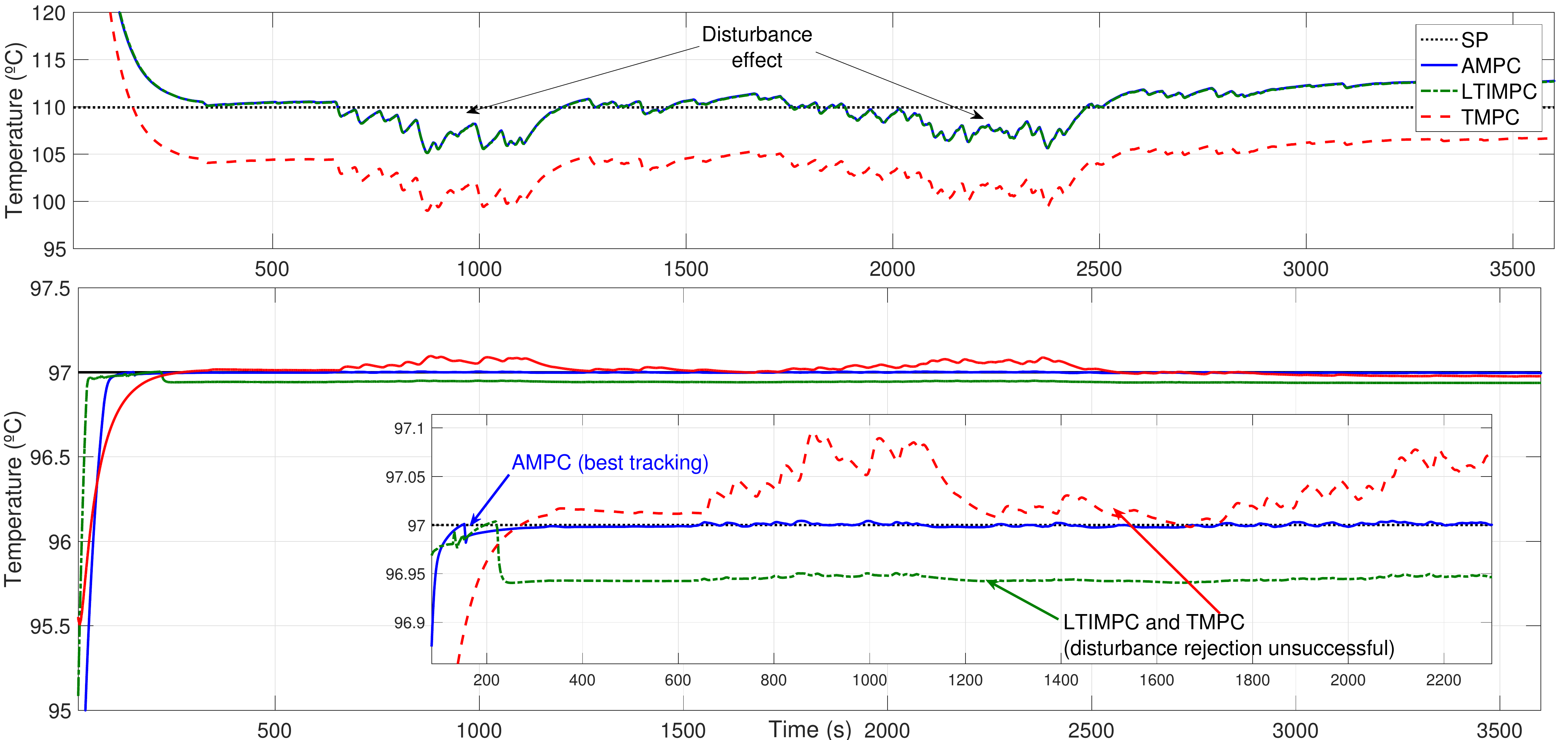}
\caption{Simulation of the Fluid and Plate Temperature Behaviours.}
\label{ResultsSTCollectoroutputs}
\end{figure}

\begin{figure}[ht]
\centering
\includegraphics[width=\linewidth]{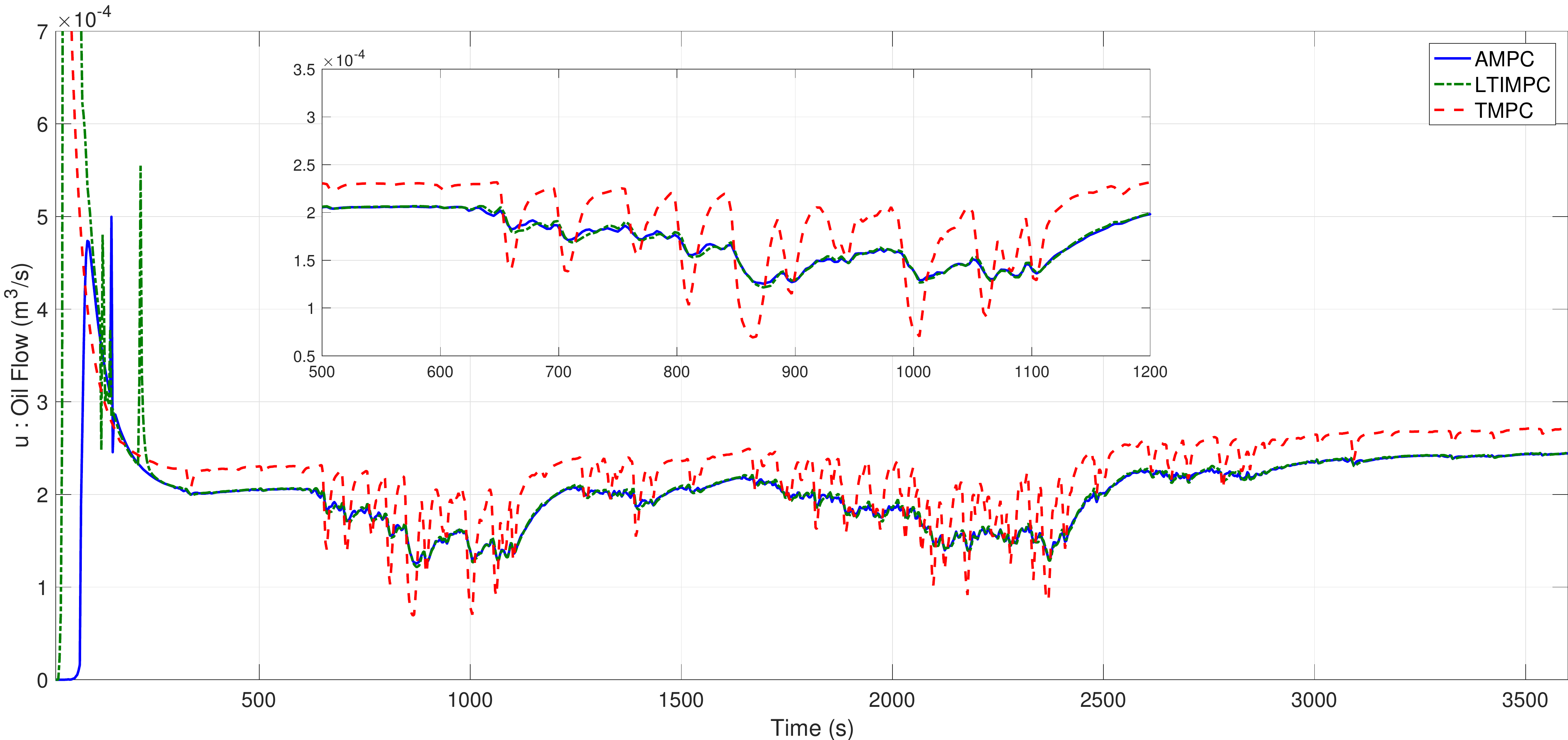}
\caption{Result of the Control Policies through the evolution of the Oil Flow signal. }
\label{ResultsSTCollectorControlSingal}
\end{figure}

Figure \ref{ResultsSTCollectoroutputs} exhibits the achieved performances for the simulation run in terms of reference tracking and disturbance rejection, showing the evolution for the fluid and plate temperatures with the three controllers. All methods guarantee regulation and disturbances-to-state stabilization. The respective applied control signal (oil flow) is given in Figure \ref{ResultsSTCollectorControlSingal}.

On one hand, it is evident that the AMPC is able to guarantee almost offset-free reference-tracking (with $\pm 0.5 \, \rm{^oC}$ tracking error) for both states, despite the abrupt solar load disturbances at $t= 700 \, \rm{s}$ and $2000 \, \rm{s}$. The LTIMPC method, as it holds an average model for prediction, can still yield roughly good tracking, but is heavily perturbed by the load disturbance (its rejection is very poor). Regarding the TMPC technique, it is able to guarantee stability, but cannot yield almost offset-free reference-tracking; it yields sufficient results, $x_1$ gets close to the set-point, but maintains non-null error regime. This is due to fact that the TMPC method uses the model in Eq. \eqref{modelLTIunc} which exhibits larger model-process mismatches, since $\xi (k)$ is bounded within a larger set. Since the AMPC method minimizes $\mu$ to find the best possible prediction model for the next $N$ steps with minimal uncertainty, it is implicitly always trying to minimize $\xi (k)$. The LTIMPC could ensure stabilization but there were no formal guarantees that it would, since the bounds model uncertainties are not taken into account (as done with the TMPC design).

To further illustrate the results, Table \ref{tabIAEs} shows the integral of the average error (IAE) index for the reference tracking and disturbance rejection results presented in Figure \ref{ResultsSTCollectoroutputs} (regarding $x_2$). Moreover, this Table also presents the Total Variance (TV) index for the three methods, which measures the total variation of the control signal over the simulation, this is:
\begin{eqnarray}
\text{TV} &:=& \sum |\delta u(k)| \, = \, \sum |u(k+1)-u(k)| \, \text{.}
\end{eqnarray}
Bigger values for the TV index shows that more variation is applied to the control along the simulation; therefore, values closer to zero indicate better (smoother) control strategies in terms of the use of the actuator.

Clearly, the best results are obtained with the AMPC method: even having the smallest TV value, which means that less control effort was necessary, the proposed tool enabled better tracking and disturbance rejection responses. The IAE indexes for this method are the ones closer the zero, which stands for perfect tracking/rejection. This is a very important issue from a practical point-of-view, since it means that the system actuators will have longer lifespan (in this case, the oil pump). These quantitative results confirm the better performance of the AMPC observed in Figures \ref{ResultsSTCollectoroutputs} and \ref{ResultsSTCollectorControlSingal}.

\begin{table}[htbp]
    \caption{\label{tabIAEs} Performance Indexes of the Control Methods.}
    \centering
	\begin{tabular}{| c | c | c | c |}
		\hline
        & \textbf{IAE Tracking} & \textbf{IAE Rejection} & \textbf{TV}\\ \hline
        AMPC & $\mathbf{0.137}$ & $\mathbf{0.0019}$ & $\mathbf{0.022} \, \rm{m^3/s}$ \\
        TMPC & $0.192$ & $0.0572$ & $0.036 \, \rm{m^3/s}$\\
        LTIMPC & $0.160$ & $0.0391$ & $0.027 \, \rm{m^3/s}$ \\ \hline
    \end{tabular}
\end{table}

\begin{figure}[ht]
\centering
\includegraphics[width=\linewidth]{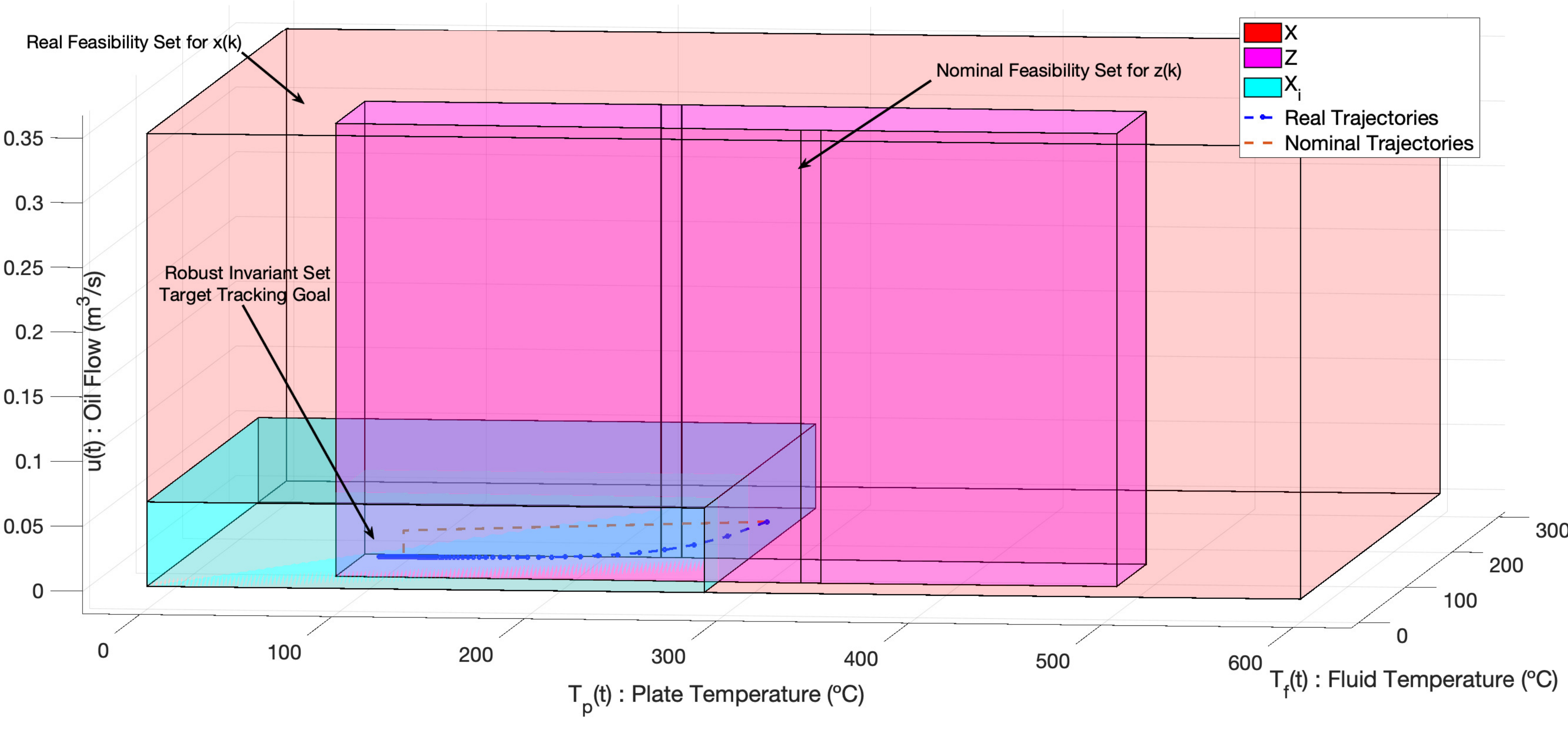}
\caption{Robust Control Sets (Nominal and Uncertain system) for which the TMPC is designed (and the respective trajectories).}
\label{ControlSetsTubes}
\end{figure}

\begin{figure}[ht]
\centering
\includegraphics[width=\linewidth]{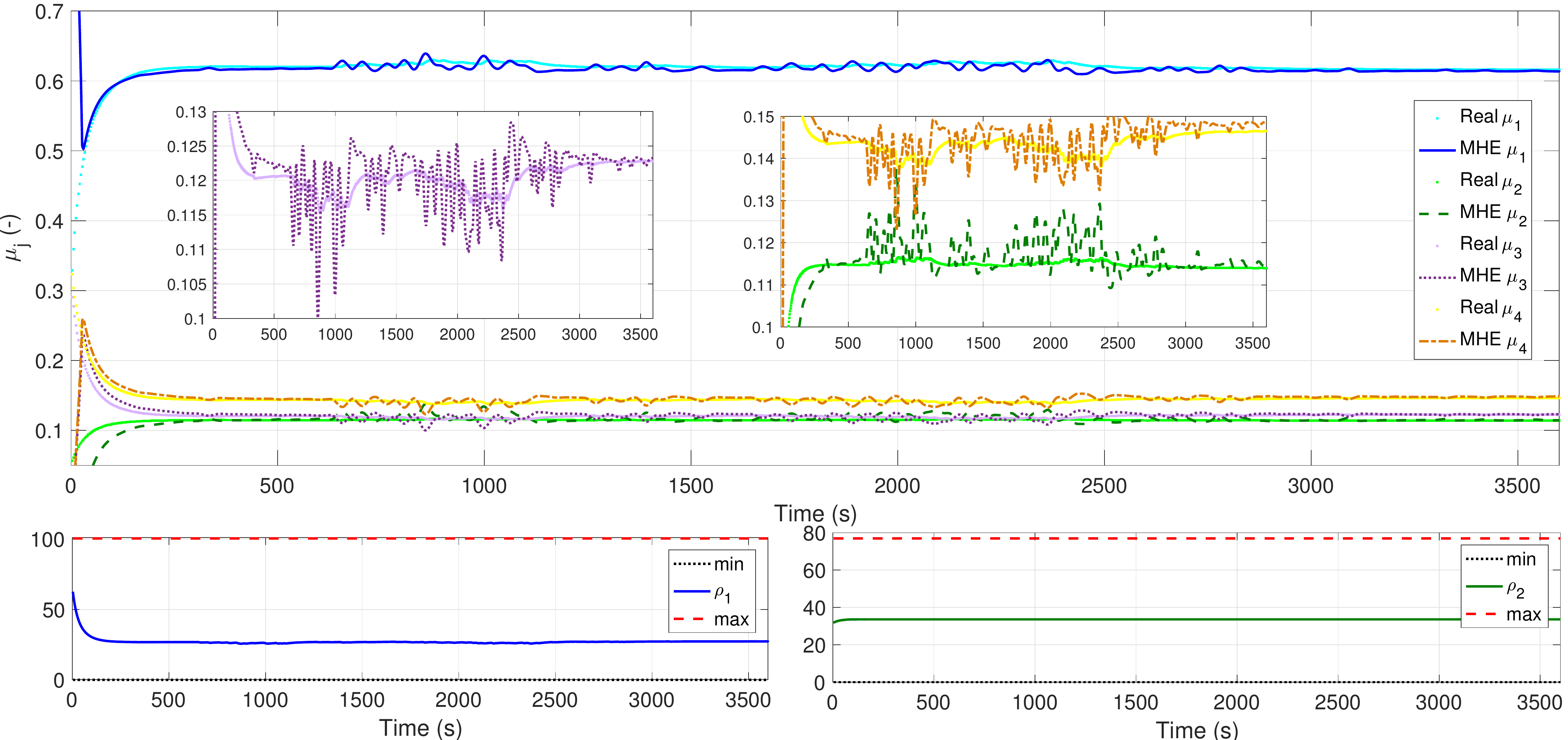}
\caption{Evolution of the polytopic membership variable (AMPC method).}
\label{membershipevolution}
\end{figure}

Figure \ref{ControlSetsTubes} shows the feasibility sets $\mathcal{X}$ and $\mathcal{Z}$ and the control invariant set $\mathbf{X}_i$; therein the systems trajectories are shown for the real states $x(k)$ and nominal states $z(k)$, departing from their initial feasibility sets to the robust invariant set that contains the tracking target. As discussed, the AMPC is able to find a better prediction model by adapting the LTI prediction model online through the polytopic variables $\mu_j$, which are shown in Figure \ref{membershipevolution}, along with the real values for $\mu_j$. This Figure also depicts the behaviour for $\rho_1(k)$ and $\rho_2(k)$, which influence $\mu_j(k)$ as given in Eq. \eqref{describemus}.

Some final considerations must be presented: 
\begin{itemize}
\item While, for the considered process, the AMPC yielded fast stabilization, formal proofs of recursive feasibility must still be provided, which shall be done in future works. This proofs are easily extended from those for the MPC for Tracking design \cite{limon2008mpc}, considering bounded disturbances due to the model-process mismatches. The TMPC has already formal stabilization and recursive feasibility proofs in the literature \cite{hanema2017stabilizing}, but did not ensure enough performances. The LTIMPC method cannot ensure stabilization or recursive feasibility, since the average-weighted LTI model may not provide enough robustness under closed-loop (i.e. for some given condition, it may run out of control).
\item Considering \textit{a-priori} preparations, the proposed AMPC does not need any procedure (as well as the LTIMPC), while the majority of RMPC methods do require them. The TMPC design, for instance, requires a planning of the nominal trajectory envelope, which can lead to quite hard offline optimization programs.
\item The number of constraints for the \textit{online} control optimization QP, considering the three methods, are similar (hard constraints on inputs and states and terminal ingredients), but the proposed AMPC has an additional parameter estimation QP, which adds additional complexity due to the model-process prediction matching framework. Anyhow, for practical purposes, the online computational stress needed to solve both problems is roughly similar (although greater with the AMPC). This issue can be critical for some real-time applications and should be taken into account by the designer.
\item In terms of \textit{online} implementation, the average computational time\footnote{In an i5 CPU@2.4 GHz (2 Cores) Macintosh with 8 GB of RAM.} elapsed to perform the three algorithms was under the sampling period of $3 \, \rm{s}$. Respectively, it took, in average, $655.5 \, \rm{ms}$ and $41.57 \, \rm{ms}$ to solve the TMPC and AMPC algorithms. The LTIMPC is solved within $41.03 \, \rm{ms}$. Note that the AMPC takes almost the same time as the LTIMPC to be computed and performs much better.
\item These results are undoubtedly interesting, since they show that the proposed AMPC algorithm can guarantee the regulation of nonlinear processes with LPV models, while using very simple mathematical tools (an LTI model and two regular constrained QPs). Moreover, it is evidenced that, for the chosen solar-thermal application, the AMPC method can outperform the robust TMPC method from the literature and well as a linearization-based MPC, which is implemented in many experimental essays \cite{ayala2011local}.
\end{itemize}

Synthetically, a new, effective QP-based adaptive method for LPV MPC design for Solar Collectors was developed in this paper. Furthermore, qualitative and quantitative discussions are presented on how to choose whether a TMPC method or a sub-optimal tool (as the proposed AMPC) is should be used for the considered nonlinear temperature tracking purposes. It is shown how the AMPC method achieves the best reference tracking performance with less total variance on the amount of oil pumped through the panels.


\section{Conclusions}\label{seconc}
This paper presented a new state-feedback MPC strategy, based on an adaptive set-based approach for systems with LPV models, for the outlet fluid temperature control of a solar-thermal heating collector, which is an inherently nonlinear process. This adaptive algorithm simplifies the nonlinear scheduling into an LTI framework using an adaptation variable for the prediction of the future system behaviour within the horizon. This adaptation variable is optimized online so to minimize model-process mismatches. In terms of the outlet fluid temperature control of this collector, the proposed method is compared to a tube-based MPC, showing enhanced performances in terms of the outlet fluid temperature control of the plant.  Some discussions were presented on how to choose which kind of method is suitable for applications with sampling time in the range of a few seconds. For further works, formal proofs of recursive feasibility will be presented for the developed method. Moreover, filtering methods (such as $H_2$ norm minimization) will be included into the MHE design procedure so that smoother estimation results for $\mu_j$ can be obtained.

\section*{Acknowledgments}
The Authors thank \textit{Universidad Tecnol\'ogica Nacional} and \emph{CNPq} under projects \textit{CNPq 401126/2014-5}, \textit{CNPq 303702/2011-7}, \textit{CNPq 305785/2015-0}, \textit{CNPq 304032/2019-0} for financing this work. 

\subsection*{Author contributions}
All authors have contributed equally for this paper.

\subsection*{Financial disclosure}
None reported.

\subsection*{Conflict of interest}
The authors declare no potential conflict of interests.


\bibliographystyle{model5-names}
\bibliography{refsplano}

\end{document}